# Optical Isolation Can Occur in Linear and Passive Silicon Photonic Structures

Chen Wang and Zhi-Yuan Li∗

Laboratory of Optical Physics, Institute of Physics, Chinese Academy of Sciences,

P. O. Box 603, Beijing 100190, China

On-chip optical isolators play a key role in optical communications and computing based on silicon integrated photonic structures. Recently there have raised great attentions and hot controversies upon isolation of light via linear and passive photonic structures. Here we analyze the optical isolation properties of a silicon photonic crystal slab heterojunction diode by comparing the forward transmissivity and round-trip reflectivity of in-plane infrared light across the structure. The round-trip reflectivity is much smaller than the forward transmissivity, justifying good isolation. The considerable effective nonreciprocal transport of in-plane signal light in the linear and passive silicon optical diode is attributed to the information dissipation and selective modal conversion in the multiple-channel structure and has no conflict with reciprocal principle.

Nonreciprocal transmission is fundamental in information processing [1]. It provides critical functionalities such as optical isolation and circulation in photonic systems. Although widely used in lasers and optical communications, such devices are still lacking in semiconductor integrated photonic systems because of challenges in both materials integration and device design [2-4]. Conventionally, the efficient routine to create nonreciprocal transmission is via time-reversal symmetry breaking [5,6], which could lead to optical isolation in devices where the forward and backward transmissivity of light is very much different. Up to now several schemes have been implemented to break reciprocity, including magneto-optical isolators [7-9], nonlinear optical structures [10,11], and time-dependent optical structures [12,13].

However, practical applications of these approaches are limited for the rapidly growing field of silicon photonics because of their incompatibility with conventional complementary metal-oxide semiconductor (CMOS) processing. Si optical chips have demonstrated integrated capabilities of generating, modulating, processing and detecting light signals for next-generation optical communications [14-16]. In 2011 several groups reported on-chip silicon diodes in the regime of time-reversal symmetry breaking. Ross's group uses traditional magnetism to construct isolators by monolithically integrating a phase-pure polycrystalline $(Ce_1Y_2)Fe_5O_{12}$ (Ce:YIG) films on silicon [17]. Their diode has good isolation signal but highly depends on the external magnetic field which could influence other devices near the diode. Qi's group reports an on-chip optical diode by using the optical nonlinearity of silicon [18]. Their diode is truly passive without external field, but has large loss of isolation signal (at least -35dB loss), a relatively large size, and slow response due to the usage of high-Q ring resonators.

Recently several schemes to realize nonreciprocal transport of light through linear and passive photonic structures have been proposed [19-21], which are essentially based on the principle of spatial-inversion symmetry breaking. In Feng *et al.* reports a passive silicon optical diode based on one-way guided mode conversion [20] and claims that nonreciprocal light propagation can be achieved in the structure. However, the result has raised hot controversies [22,23]. S. Fan *et al.* assess the reciprocity by deriving and analyzing the corresponding scattering matrix for relevant forward and backward modes of the structure [22]. They argue that the structure cannot enable optical isolation with nonreciprocal light transport because it possesses a symmetric coupling scattering matrix. In their response, Feng *et al.* acknowledge that their structure, as a one-way mode converter with asymmetric mode conversion, is Lorentz reciprocal and on its own cannot be used as the basis of an optical isolator. The controversies have thus raised a fundamental question: Can one construct an optical isolator by using a linear and time-independent optical system? The answer to this question by the authors of Ref. [22,23] obviously is no.

In December 2011, we reported an ultrasmall on-chip optical diode based on silicon photonic crystal slab heterojunction structures [21]. The optical diode is linear, passive, and time-independent, but has a spatial-inversion symmetry breaking geometry. Our numerical calculations and experimental measurements both show that the forward and backward transmission efficiencies of the diode are very much different at the wavelengths around 1,550 nm, with a signal contrast reaching a high value of about 89%. It is the central issue of the current paper to further discuss the optical isolation performance of this structure and address the fundamental physics issues raised in the context of current controversies [22,23]. Our numerical simulations and scattering matrix analyses both show that the diode can construct an optical isolator in no conflict with any reciprocal principle.

Our optical diodes are made from the heterojunction between two different silicon two-dimensional square-lattice photonic crystal slabs with directional bandgap mismatch and different mode transitions [21]. To facilitate the convenience of discussion, we reproduce the geometry of the diode structures and corresponding numerical and experimental data in Fig. 1. The diode patterns were first defined in resist using the electron beam lithography on the top layer of a silicon-on-insulator chip. The resist patterns were then transferred to silicon layer using the inductive coupled plasma reactive ion etching technique. The lattice constant $a$ was set to 440 nm, the two radii $r_1$ and $r_2$ were approximately equal to 110 nm and 160 nm, respectively, and the slab thickness was 220 nm. The insulator layer (SiO$_2$) underneath the silicon pattern regions was finally removed by a HF solution to form an air-bridged structure. The experimental forward transmissivity approaches 21.3% and the best signal contrast $S$ of the diode structure reaches 0.885 at the peak, which is near the value of the present electrical diodes. The overall size of the ultrasmall diode is $6 \times 6 \mu m^2$. Both theory and experiment show an obvious significant unidirectional infrared light transport in the silicon diode without time-reversal symmetry breaking.

A question naturally arises: Is there a good isolation effect of the silicon diodes? To answer this question, we implement a direct method, which is to put a total reflection mirror after the output port in the forward direction and monitor the reflection signal from the input port (which records the round-trip reflectivity). If the reflection signal is the same as or comparable with the forward signal, then the structure does not have the isolation property. In contrast, if the reflection signal is much smaller than the forward signal, then a good isolation property is implied. This definition is similar to the case in conventional magneto-optical isolators used in lasers. An equivalent way to investigate the structure with the total reflection mirror is to adopt a doubled structure with a symmetrical plane at the output port in the forward direction, as depicted in Fig. 2. This method has the meaning of putting the forward output as the backward input to test the isolation property of the structure. By implementing this method, we calculate simultaneously the forward transmissivity and the round-trip reflectivity of the two diode structures as displayed in Fig. 1 by using three-dimensional finite-difference time-domain (3D-FDTD) method. Comparison of these two quantities would directly measure their isolation properties.

Figure 2(a) is the schematic geometry of the doubled-diode structure corresponding to the diode depicted in Fig. 1(a). The parameters of the diode are the same as in Fig. 1. The width of the input, output and reflection waveguides is the same as $2a$ ($a$=440nm). The spectrum on the round-trip reflection port [Fig. 2(b)] shows that the reflection peak is located at 1,553 nm and the maximum reflectivity is only 0.9%, while the maximum transmissivity of the forward peak at 1,553 nm is 12.8% [shown in Fig. 1(b)], which is nearly 15 times larger than the reflection signal. The fact that the round-trip reflectivity is more than one order of magnitude smaller than the forward transmissivity clearly shows that the diode structure has a good isolation property.

We construct another doubled-diode structure based on the diode structure illustrated in Fig. 1(d) and calculate its isolation property. The width of the input and reflection waveguides is the same $2a$ ($a$=440nm) and the output waveguide is $6a$. The spectra [Fig. 2(d)] show that the reflection peak is located at 1,582 nm and has a round-trip reflectivity of only 0.3%, which is almost two orders of magnitude smaller than the forward peak [with a maximum transmissivity of 22.9%, shown in Fig. 1(d)].

The result indicates that the diode has a better isolation property.

Another way proposed to break spatial-inversion symmetry is using photonic crystal gratings [19]. Former study has shown that the grating structure can cause the one-way transmission effect in two-dimensional system, which could be used for designing optical diodes [19]. Also, the same question is asked: Can this grating structure be used to construct a good isolator? To have a definite answer to this question, we also double this grating structure to test its isolation property. The schematic geometry of doubled-grating structure is illustrated in Fig. 3(a), from which the round-trip reflection spectra for the grating can be calculated. In addition, the schematic geometry of the single grating structure used to calculate the forward and backward transmission spectra is depicted in Fig. 3(b). The lattice constant $a$ of the grating was set to be 490 nm, and the radius $r$ is $r=0.2a$ (the same as in Ref. 19), so that the one-way transport wavelengths are located around 1,550 nm. The calculated forward transmission, backward transmission, and the round-trip reflection spectra of the photonic crystal grating structure are displayed in Fig. 3(c). The results show a remarkable behavior of unidirectional transmission between 1,580 nm and 1,620 nm [see the black line and blue line in Fig. 3(c)], where the backward transmission is much lower than the forward transmission. However, the round-trip reflection signal in this frequency region is nearly the same as the forward transmission signal [compare the black line and red line in Fig. 3(c)], indicating that the unidirectional-transport photonic crystal grating does not have the true isolation property.

To better understand the fundamental physics, we further show via a detailed analyses based on the scattering matrix theory adopted in Ref. [22,23] that the above numerical results of optical isolation are in no conflict with the reciprocity theorem involved in our linear and passive silicon optical diode structure. As is depicted in Fig. 4, our diode basically consists of two in-plane information channels (A and B, the input and output waveguide channels for infrared signal, which can be either single mode or multimode channels.) as well as many in-plane and off-plane scattering channels (denoted as C as a whole, which causes dissipation of information away the

signal channels). At the two ends of the diode device the fields are written as follows:

$$\begin{bmatrix} A_{out} \\ B_{out} \\ C_{out} \end{bmatrix} = S \begin{bmatrix} A_{in} \\ B_{in} \\ C_{in} \end{bmatrix}, \quad (1)$$

in which $A_{in}$ corresponds to the input signal from port A, $A_{out}$ to the output signal from port A, $B_{in}$ to the input signal from port B, $B_{out}$ to the output signal from port B, $C_{in}$ to the input signal from port C, and $C_{out}$ to the output signal from port C. The scattering matrix $S$ transforms the input state of all the channels [the column vector in the right hand of Eq. (1)] into the output state of all the channels [the column vector in the left hand of Eq. (1)]. In the case of forward transport of infrared signal across the diode with an input signal $a_{in0}$ at port A, the reflection signal $a_{out1}$ is very small compared with the transmission signal $b_{out1}$ or the scattering signal $c_{out1}$ according to our numerical simulation results [21]. Then the scattering equation of the forward transmission is written as:

$$\begin{bmatrix} 0 \\ b_{out1} \\ c_{out1} \end{bmatrix} = S \begin{bmatrix} a_{in0} \\ 0 \\ 0 \end{bmatrix}. \quad (2)$$

As the silicon diode structure is linear and passive, the system as a whole is reciprocal in regard to time-reversal symmetry, so the scattering matrix $S$ is symmetric and satisfies [22-26]:

$$S = S^{-1}. \quad (3)$$

Suppose all the output signals are reversed and come back into the system, then the input at port B for the system is now exactly the same as $\begin{bmatrix} 0 & b_{out1} & c_{out1} \end{bmatrix}^T$. The scattering equation is then

$$S \begin{bmatrix} 0 \\ b_{out1} \\ c_{out1} \end{bmatrix} = S^{-1} \begin{bmatrix} 0 \\ b_{out1} \\ c_{out1} \end{bmatrix} = \begin{bmatrix} a_{in0} \\ 0 \\ 0 \end{bmatrix}, \quad (4)$$

which is exactly the same as the initial input from port A. This clearly indicates that there is no nonreciprocal transmission behavior in the structure if all information is reversed back into the system, consistent with the reciprocity theorem for a time-reversal symmetric system.

However, the story can be very different when the in-plane signal transport is concerned. In our structure the information and energy involved in C channels are dissipated permanently against the in-plane channel A and B due to scattering loss (both in-plane and off-plane), and they cannot be reversed back totally and input again into the structure, so in practice, $C_{in}$ in Eq. (1) can be assumed to be zero. As a result, Eq. (4) should be modified as:

$$S \begin{bmatrix} 0 \\ b_{out1} \\ 0 \end{bmatrix} = \begin{bmatrix} a_{out2} \\ b_{out2} \\ c_{out2} \end{bmatrix}. \tag{5}$$

In general, Eq. (5) looks very different from Eq. (4), which indicates that the reciprocal transport of light in regard to the signal channel A and B has been broken. It shows that even if the same forward transmission signal of port B is reversed back and input into the diode, the output signal of port A can be much different from the initial input signal $a_{in0}$ of port A because no signal is reversed and input back into the channel C. Therefore, the considerable nonreciprocal transmission behavior can take place for the in-plane signal with no conflict with the reciprocal principle. According to our simulations and experiments for the silicon optical diode, $a_{out2}$ and $b_{out2}$ are much smaller than $a_{in0}$. This should justify that a good isolation effect can occur in the silicon optical diode. In ideal structures, both of them are zero, and Eq. (5) becomes

$$S \begin{bmatrix} 0 \\ b_{out1} \\ 0 \end{bmatrix} = \begin{bmatrix} 0 \\ 0 \\ c_{out2} \end{bmatrix}, \tag{6}$$

which implies a 100% signal contrast of the isolator.

The most remarkable difference between our diode design in Ref. [21] and the asymmetric mode-conversion waveguide diode structure in Ref. [20] is that the latter structure has only two channels: one input and one output channel (each involving one even and one odd mode). As all light signals are contained within the channels during the transport, scattering and modal conversion process across the diode structure, as well as the reflection process with respect to a mirror and the signal dissipation to other channels does not occur, the mode conversion of the system must be reciprocal under the circumstance of the symmetric scattering matrix. In contrast, our diode has not only two in-plane signal channels but also many other scattering channels. In this linear and passive structure, the working channels are only two selected channels among the multiple channels, and the other unselected channels can help the structure break the spatial inversion symmetry without changing the symmetric scattering matrix circumstance as these channels cannot reverse the output signals back into the structure. As a result, significant nonreciprocal transport of light can occur in the signal channels in no conflict with reciprocal principle. Simply speaking, it is the selective mode conversion in a multiple-channel structure that comprises the basis of optical isolation in our passive, linear, and time-independent silicon optical diode. The above picture can also help to explain the bad isolation property of the photonic crystal grating in Fig. 3. Although the structure itself involves multiple channels of signal, however, all output signals are reversed and input back into the grating itself by the mirror.

It is worth saying a few more words here for better drawing a clear picture about the physics discussed in the above. In nature, as time always flows forward and cannot be reversed, one usually uses the term of reciprocal or nonreciprocal transport of light to describe a model system of back transport of light, in many cases to describe the reflection of light back into the considered structure. In this regard, simply consider a point source radiating an outgoing spherical wave front. If time can be reversed, the outgoing spherical wave front is contracted into an ingoing spherical wave front, eventually to a point. This is a very good picture to describe reciprocal transport of light in a linear system. However, to realize in real world such a concept, one needs to place a perfect spherical mirror concentric with the point source of light, which reflects back all information carried by the outgoing expanding spherical wave into the ingoing contracting spherical wave. If, however, one has only a small planar mirror placed at some distance and with a limited solid angle with respect to the source, the reflected signal can never return to the initial state of a point source when it reaches the position where the light source is located. The conventional magneto-optical isolator also works in this category of physical picture. It is used to block down the back-reflection signal of the transmission light, and the underlying physics can be well described by the model of time-reversal symmetry breaking. The same physics picture applies equally well to our optical diode. The fact that there exists information dissipation from the signal channels to other channels in a spatial-inversion symmetry breaking structure is sufficient to induce an effective nonreciprocal transport and optical isolation in regard to the signal.

In summary, we have analyzed the optical isolation properties of a silicon photonic crystal slab heterojunction diode by comparing the forward transmissivity and round-trip reflectivity of in-plane infrared light across the structure, aiming to address the fundamental issues that have been raised in the great attentions and hot controversies upon the nonreciprocal transport of light through linear and passive photonic structures. Our numerical calculations show that the round-trip reflectivity is much smaller than the forward transmissivity, and this justifies that good isolation does take place in this silicon diode structure. Our scattering matrix analysis indicates that the considerable effective nonreciprocal transport of in-plane signal light can be attributed to the information dissipation and selective modal conversion in the multiple-channel spatial-inversion symmetry breaking structure and has no conflict with reciprocal principle for a time-reversal symmetric structure. That optical isolation can occur in a linear, passive, and time-independent optical structure would stimulate more thinking on the general transport theory of light in the fundamental side and open up a road towards photonic logics in silicon integrated optical devices and circuits in the application side.

This work was supported by the National Basic Research Foundation of China under grant no. 2011CB922002 and Knowledge Innovation Program of the Chinese Academy of Sciences (No. Y1V2013L11) .

∗To whom correspondence should be addressed.

lizy@aphy.iphy.ac.cn

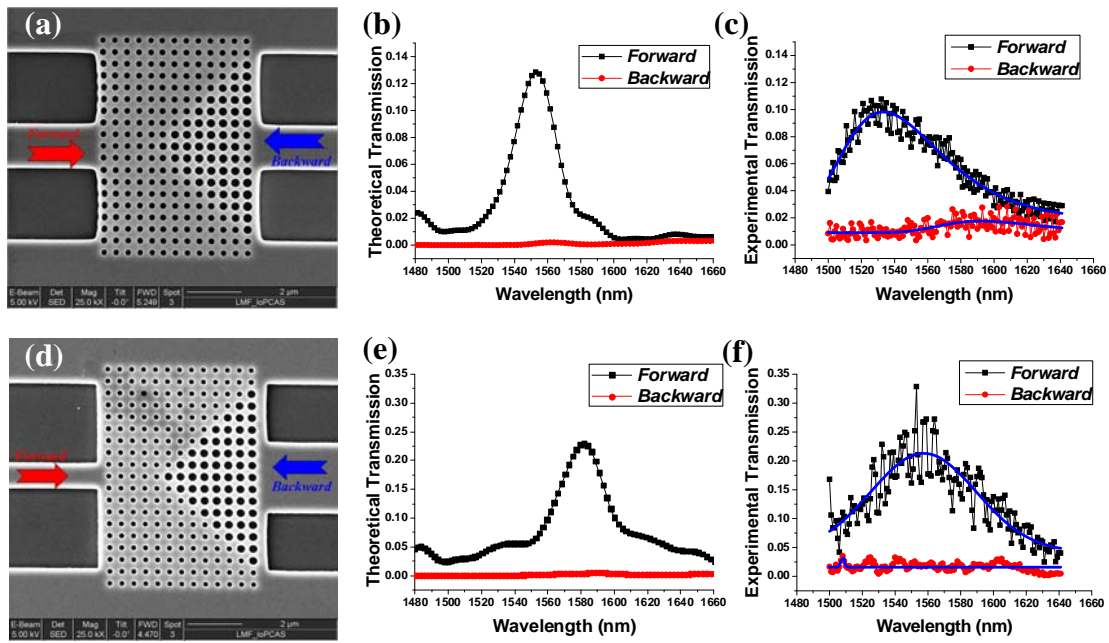

Fig. 1. (Color online) (a) Scanning electron microscope images, and (b) theoretical and (c) experimental transmission spectra of an optical diode structure. (d) Scanning electron microscope images, and (e) theoretical and (f) experimental transmission spectra of another optical diode structure.

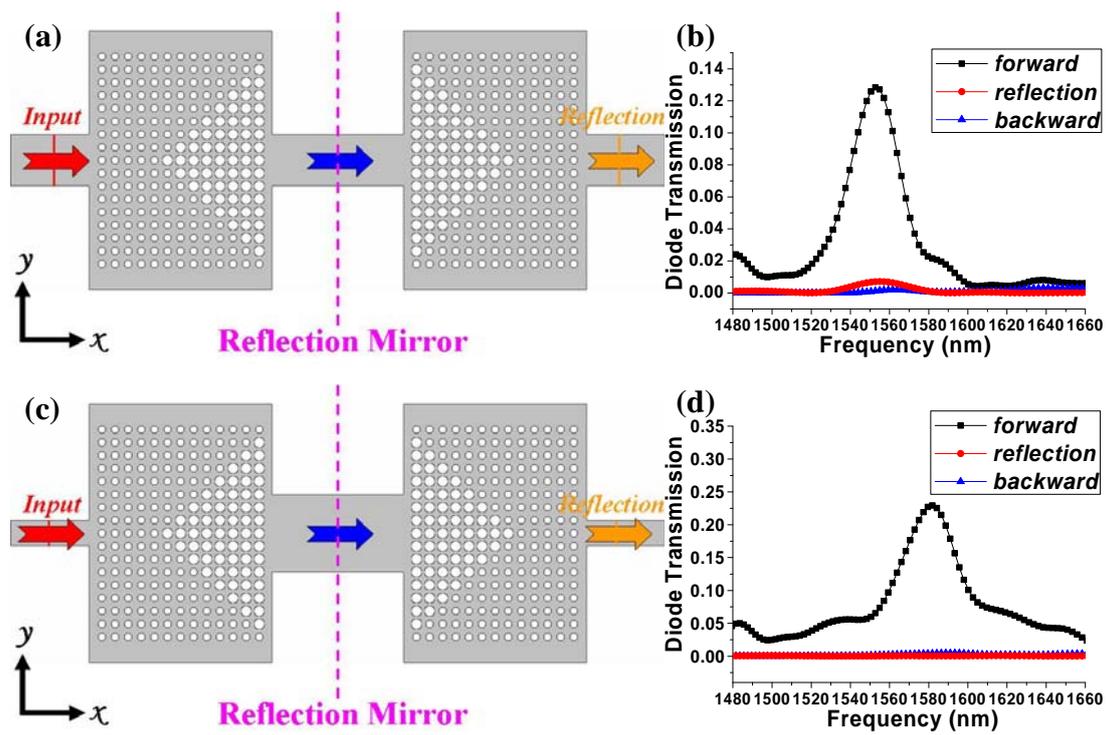

Fig. 2. (Color online) (a)(c) Schematic geometry of our doubled-diode structure with the total reflection mirror. (b)(d) Simulated transmission spectra of the diode in forward transmission (black line), backward transmission (blue line) [shown in Fig. 1], and the round-trip reflection (red line).

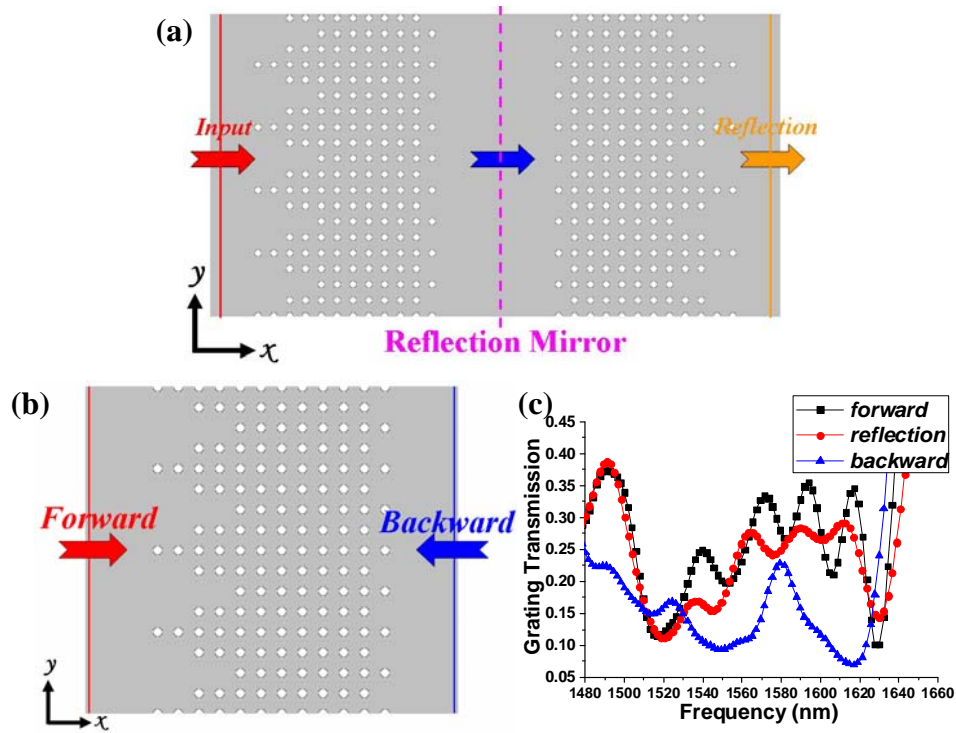

Fig. 3. (Color online) (a) Schematic geometry of (a) a doubled-grating structure with the total reflection mirror, and (b) the corresponding single-grating structure under forward and backward transmissions. (c) Simulated forward transmission (black line), backward transmission (blue line), and the round-trip reflection (red line) spectra of the grating.

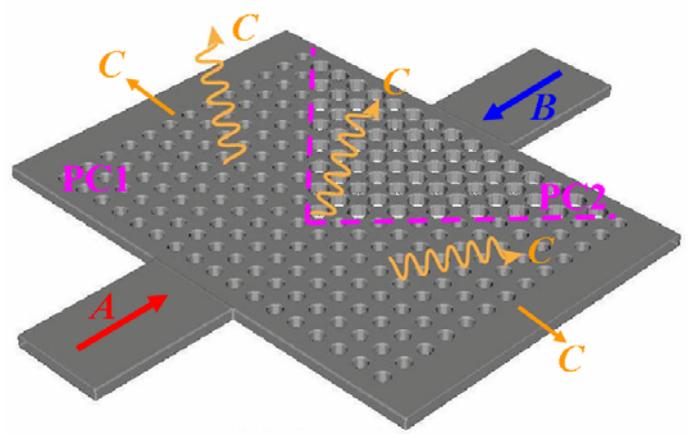

Fig. 4. (Color online) Schematic geometry of the diode structure as shown in Fig. 1 for multiple-channel mode scattering and conversion analysis.